\documentclass[10pt,conference]{IEEEtran}
\usepackage{cite}
\usepackage[dvips]{color}
\usepackage{tabu}
\usepackage{comment}
\usepackage{todonotes}
\usepackage{epsf}
\usepackage{epsfig}
\usepackage{times}
\usepackage{epsfig}
\usepackage{graphicx}

\usepackage{mathtools}
\usepackage{mathrsfs}
\usepackage{amssymb,amsmath}
\usepackage{amsmath}
\usepackage{amsfonts}

\usepackage{algorithmicx}  
\usepackage[algo2e]{algorithm2e} 
\usepackage{cleveref}
\usepackage{dsfont}
\usepackage{lettrine} 
\usepackage{epsfig,algorithm,algpseudocode,amsthm,url}
\usepackage{caption}
\usepackage{subcaption}
\usepackage[justification=raggedright]{caption}
\usepackage{subcaption}
\usepackage{bbm}
\usepackage[utf8]{inputenc}
\usepackage{verbatim}
\usepackage{setspace}	
\usepackage[T1]{fontenc}
\usepackage{textcomp}
\usepackage{tabularx}
\usepackage{xpatch}
\DeclareMathOperator*{\argmax}{argmax}

\usepackage{stackengine}
\usepackage[nocomma]{optidef}
\usepackage{commath}

\usepackage[left=1.62cm,right=1.62cm,top=1.85cm,bottom=2.6cm]{geometry}

\IEEEoverridecommandlockouts 

\begin{document}
\title{\Huge Attention-based Open RAN Slice Management using Deep Reinforcement Learning \thanks{This material is based upon work supported by the Air Force Office of Scientific Research under award number FA9550-20-1-0090 and the National Science Foundation under Grant Numbers CNS-2232048, and CNS-2204445.} \thanks{DISTRIBUTION STATEMENT A: Approved for Public Release; distribution unlimited AFRL-2023-0508 on 6 June 2023.}


}
\author{
	\IEEEauthorblockN{
	Fatemeh Lotfi\IEEEauthorrefmark{1}, Fatemeh Afghah\IEEEauthorrefmark{1}, Jonathan Ashdown \IEEEauthorrefmark{2}}

	\IEEEauthorblockA{\IEEEauthorrefmark{1}Holcombe Department of Electrical and Computer Engineering, Clemson University, Clemson, SC, USA \\
Emails: flotfi@clemson.edu,   fafghah@clemson.edu}
\IEEEauthorblockA{\IEEEauthorrefmark{2}Air Force Research Laboratory,  Rome, NY 13441, USA \\
	 Email: jonathan.ashdown@us.af.mil}}\vspace{-0.2cm}
\maketitle\vspace{-0.3cm}
\begin{abstract}


As emerging networks such as Open Radio Access Networks (O-RAN) and 5G continue to grow, the demand for various services with different requirements is increasing. Network slicing has emerged as a potential solution to address the different service requirements. However, managing network slices while maintaining quality of services (QoS) in dynamic environments is a challenging task. Utilizing machine learning (ML) approaches for optimal control of dynamic networks can enhance network performance by preventing Service Level Agreement (SLA) violations. This is critical for dependable decision-making and satisfying the needs of emerging networks. Although RL-based control methods are effective for real-time monitoring and controlling network QoS, generalization is necessary to improve decision-making reliability. This paper introduces an innovative attention-based deep RL (ADRL) technique that leverages the O-RAN disaggregated modules and distributed agent cooperation to achieve better performance through effective information extraction and implementing generalization. The proposed method introduces a value-attention network between distributed agents to enable reliable and optimal decision-making. Simulation results demonstrate significant improvements in network performance compared to other DRL baseline methods. 
\vspace{-0.2cm}


\vspace{-0.cm}
\end{abstract}
\section{Introduction} \vspace{-0cm}
With emerging new applications such as augmented and virtual reality (AR/VR), the Internet of Everything, drone-based surveillance, and package delivery, the demand for heterogeneous services is growing every day~\cite{li2020ran}. 
These diverse use cases need a wide range of service requirements that vary in terms of throughput, coverage, traffic capacity, user density, latency, reliability, and availability. 
To meet these rigorous requirements, the development of network slicing is being explored as a solution to maintain the network's quality of service (QoS) in the face of dynamic changes and heterogeneous requirements~\cite{3gpprelease16,popovski20185g}. 

Network slicing allows users to specify their service requirements in a service level agreement (SLA), and the network operator creates customized network slices to meet those needs while maintaining isolation. This concept is essential for both core network (CN) and radio access network (RAN), where Software-Defined Network (SDN) and Network Function Virtualization (NFV) enables slicing in the CN, and radio resource management (RRM) manages physical radio resources for SLA compliance~\cite{yang2020data}. However, resource allocation in 5G RAN slicing remains challenging due to varying traffic demands, leading to underutilization or overload of certain slices, compromising QoS. To address this, a new open RAN (O-RAN) architecture has been introduced, disaggregating RAN functions into different units and leveraging machine learning (ML) for improved performance. 
O-RAN includes open central units (O-CU), open distributed units (O-DU), and open radio units (O-RU) connected to RAN's intelligent controllers (RIC) for real-time and non-real-time network management~\cite{3gpp2017study,oranslice2020,polese2022understanding}.

Network slicing can benefit from the application of artificial intelligence (AI) and ML approaches
. However, some of these methods encounter challenges such as the need for a large volume of diverse data and extensive exploration to expedite convergence and train an ML model that can generalize to various situations without affecting actual RAN performance~\cite{thaliath2022predictive,cheng2022reinforcement,zhang2022team,zhang2022federated,lotfi2022evolutionary}. Literature has recently studied O-RAN control and managing challenges~\cite{polese2022understanding,thaliath2022predictive,cheng2022reinforcement,zhang2022team,zhang2022federated,lotfi2022evolutionary}. To prevent SLA violation, \cite{thaliath2022predictive} introduced a closed-loop resource allocation for network slicing in the O-RAN framework. 
In~\cite{thaliath2022predictive}, a regression-based learning approach with an LSTM model is used to predict network traffic for the next hour, allowing for resource allocation optimizations. However, this offline model lacks real-time monitoring capability, leading to SLA violations. To address this, RL techniques have shown great potential in achieving optimal control of real-time dynamic network slicing. In~\cite{cheng2022reinforcement}, authors introduce an RL-based algorithm that dynamically allocates resources to a target network slice without affecting others. They utilize Q-learning with a shared Q-table between O-DU and O-CU modules for resource allocation optimization in a topology with a single O-CU and multiple O-DUs. 

In dynamic and heterogeneous networks, 
a single RL agent may not be sufficient for effectively managing the network. To address this, researchers have explored methods for enhancing network controllers' performance. In~\cite{zhang2022team}, authors propose a cooperative approach between two separate network controllers to maximize system throughput. Actions of each controller agent are defined as states of the other agent to ensure cooperation and prevent conflicts. In~\cite{zhang2022federated}, a federated RL method is presented that coordinates multiple network controllers and enhances their learning efficiency. Deep Q-learning is used for each controller, and the federated learning process combines all Q-tables to create a global model for action selection. Additionally, in~\cite{lotfi2022evolutionary}, authors propose a hybrid network slicing strategy that combines evolutionary algorithms (EA) and off-policy deep RL (DRL) to create an EDRL algorithm for O-RAN architecture. Multi-actor modules are used for the EA component and an actor-critic module for the DRL component, enabling resource allocation decisions across multiple slices. 

Despite the studies conducted in~\cite{cheng2022reinforcement, zhang2022team, zhang2022federated, lotfi2022evolutionary} to address real-time closed-loop dynamic network slicing, many of these methods lack consideration of generalization during training, which may lead to sub-optimal results in unforeseen scenarios. Additionally, these approaches often utilize all available information from collaborators without accounting for their relative importance in conveying information for agent training. As a result, there is a need for novel solutions that can effectively handle the challenge of exploring dynamic wireless networks using DRL approaches in the O-RAN slicing scenario, ensuring broad applicability and efficacy.
The main contribution of this paper is to exploit the opportunity provided by the O-RAN architecture to generate new experiences using disaggregated modules. To achieve this goal, the study adopts a distributed DRL approach where the distributed agents are located in O-DU locations and are exposed to diverse network conditions. Different exploration policies of each disaggregated module improve robustness and stabilize the learning process. However, not all of the information collected from different locations is equally important. 
To address this issue, in this paper, we formulate a problem aimed at minimizing the SLA violation in a wireless communication network while taking into account resource constraints. To solve this problem, we propose a novel attention-based DRL (ADRL) strategy for cooperation between distributed agents. The proposed ADRL algorithm models the O-RAN slicing management problem as a Markov decision process (MDP) and introduces a value attention network that is applied to distributed agents embedding samples. 
By using this network, a global critic network can identify which part of the network requires more attention to have reliable decision-making. The proposed ADRL algorithm is evaluated through simulations, and the results demonstrate a $32.8\%$ improvement in the final return value of the intelligent network management compared to the distributed-DRL baseline method. \emph{To the best of our knowledge, this is the first work that utilizes an attention-based DRL method for network slicing in O-RANs}. 

The rest of this paper is organized as follows. Section \ref{sysmodl} presents the system model and the problem formulation for the O-RAN slicing. Section \ref{ADRL} presents the proposed attention-based DRL method and the ADRL algorithm to solve the MDP problem. Simulation results are presented in Sec. \ref{sim} and conclusions are drawn in Sec. \ref{conclusion}.\vspace{-0.2cm}

\begin{figure}[t!]
  \centering
    \includegraphics[width=0.76\columnwidth]
    {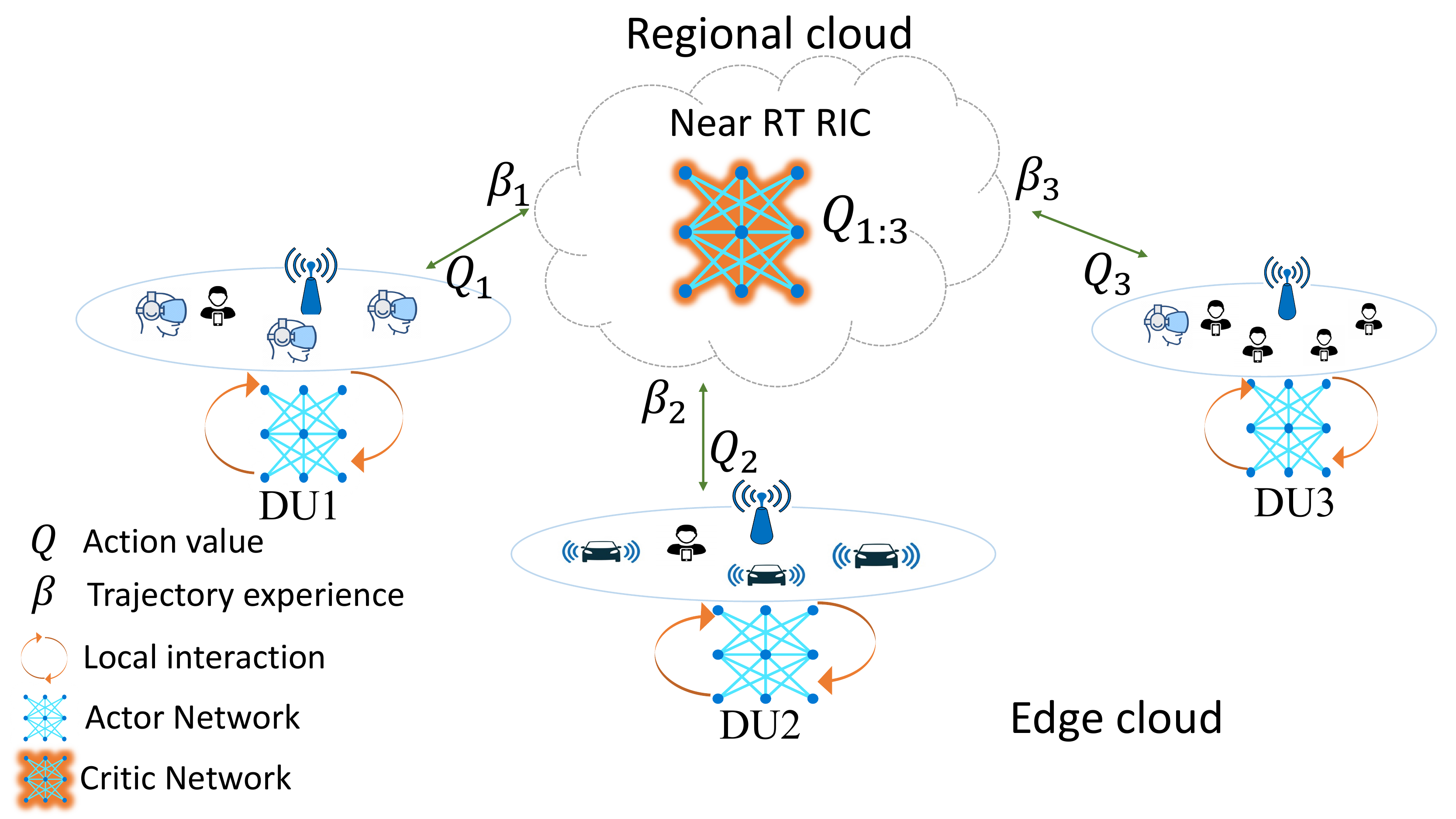}\vspace{-0.1cm}
    \caption{\small System model topology for intelligent O-RAN slice management. 
    }\vspace{-0.2cm}
    \label{sys_graph}
\end{figure}
\section{System model}\label{sysmodl}
Consider an O-RAN based network structure with $L=3$ types of slides in a set $\mathcal{L}$ for downlink transmission. O-RAN system is composed of multiple CUs, DUs, and two RICs as a non-real time RIC and a near-real time RIC. In this network, $N$ heterogeneous users in a set $\mathcal{N}$ served by different network slices with specific QoS demands as $Q_l, l \in \mathcal{L}$. Each slice $l \in \mathcal{L}$ serves $N_l$ user equipment (UE) with similar QoS. Additionally, to ensure that the QoS requirements for each assigned UE are met, the slices must have access to a common set of limited resources, with a total amount of $K$. The network goal is to satisfy these diverse QoS demands of slices by intelligently controlling resources between slices in the RIC module. Then, the medium access control (MAC) layer must assign resources to UE devices based on the resource allocation strategy outlined by the slice management. To tackle this, we will introduce a wireless model and formulate the slice management problem with consideration to wireless resource constraints.\vspace{-0.1cm}

\subsection{Wireless communication model }
The O-RAN architecture shown in Fig. \ref{sys_graph}, employs different network slices with varying QoS criteria to serve different UEs. The RIC module manages these slices and their resources, utilizing dynamic resource management to adapt to changing wireless channels. However, this dynamic approach also complicates resource assignment due to its stochastic nature.

The QoS criteria $Q_l$ for $\mathcal{L}$ slices can be defined in terms of throughput, capacity, and latency.
The achievable QoS for slice $l$ is defined based on orthogonal frequency-division multiple access (OFDMA) schemes. The data rate for slice $l$ can be expressed as:\vspace{-0.2cm}
\begin{align}\label{urate}
    c_{l} = \lim_{\tau \to \infty} \frac{B}{\tau} \sum_{t=1}^{\tau}&\sum_{n=1}^{N_l}\sum_{k=1}^{K} e_{n,k} b_{l,k}\nonumber\\
    &\times\log\Big(1+\frac{p_u d_{n}(t)^{-\eta} |h_{n,k}(t)|^2}{ I_{n,k}(t)+ \sigma^2}\Big),
\end{align}
where $e_{n,k} \in \{0, 1\}$ is a binary variable indicating RB allocation for user $n$ in RB $k$, and $b_{l,k} \in \{0, 1\}$ is the RB allocation indicator for slice $l$ in RB $k$. $B$ represents RB bandwidth, $K$ represents total available RBs, $p_u$ denotes transmit power per RB of O-RU, and $d_n(t)$ represents the distance between user $n$ and its assigned O-RU. $\eta$ shows the path loss exponent, and $|h_{n,k}(t)|^2$ indicates the time-varying Rayleigh fading channel gain. In equation \eqref{urate}, $I_{n,k}(t)$ denotes downlink interference from neighboring O-RUs on RB $k$, and $\sigma^2$ represents the variance of additive white Gaussian noise (AWGN).

\subsection{Problem formulation}
Our goal is to minimize SLA violations across the network. To achieve this, we define a probability function that quantifies SLA violation using a threshold value $\lambda_l$ and a marginal value $\epsilon_l$, denoted as $\mathbbm{P}(\abs{Q_l(\boldsymbol{e},\boldsymbol{b})-\lambda_l} \geq \epsilon_l)$ for all slices $l$ in $\mathcal{L}$. Here, $\boldsymbol{e}$ and $\boldsymbol{b}$ represent resource allocation indicator vectors for each UE and each slice, respectively, denoted as $e_{n,k}$ and $b_{l,k}$. Accordingly, we have formulated an optimization problem that aims to identify an optimal resource allocation policy, taking into account the constraints on the total required resources. \vspace{-0.1cm}
\begin{subequations} 
\begin{align}\label{opt1}
 \argmax_{\boldsymbol{b},\boldsymbol{e}} & \hspace{0.5cm} 
 \mathbbm{P}(\abs{Q_l(\boldsymbol{e},\boldsymbol{b})-\lambda_l} \leq  \epsilon_l),\\
 \text{s.t.,} 
& \hspace{0.5cm} \forall l \in \mathcal{L},\,\,  \forall n \in \mathcal{N} , \label{opt1_q}\\
& \hspace{0.5cm}  \sum_{l=1}^{L}\sum_{n=1}^{N_l}\sum_{k=1}^{K} b_{l,k}e_{n,k} \leq K, \label{opt1_Ns}\\
& \hspace{0.5cm} \sum_{l}b_{l,k} \leq 1,\,\,   \label{opt1_e}
\end{align}
\end{subequations}\vspace{-0.1cm}
where $\mathbbm{P}$ shows a 
probability function. The QoS requirements of slice $l$ are described by the values $\lambda_l$ and $\epsilon_l$, which respectively represent the desired threshold and margin. 
Constraint \eqref{opt1_Ns} and \eqref{opt1_e} represent the RB availability limitation on allocated resources to slices and UEs. 
Problem \eqref{opt1} aims to achieve the demanded QoS for $\mathcal{L}$ slices by minimizing the probability of SLA violation. However, this problem is known to be NP-hard and involves mixed-integer stochastic optimization, making it challenging to solve. In this context, the Markov decision process (MDP) framework provides a mathematical approach for decision-making and optimization in situations involving partial randomness. Therefore, modeling \eqref{opt1} as an MDP and employing dynamic methods such as DRL approaches can be advantageous for solving this complex problem.\vspace{-0.1cm}

\subsection{MDP-based optimization problem}
We consider an intelligent agent containing the xApp located in the near-RT RIC module that makes decisions about the related environment to manage the O-RAN slicing, as shown in Fig \ref{sys_graph}. Thus, the problem can be represented as an MDP with tuples $\langle \mathcal{S},\mathcal{A}, T,\gamma,r \rangle$, where $\mathcal{S}$, $\mathcal{A}$, and $T$ represent the state space, action space, and transition probability from the current state to the next state as $P(s_{t+1}|s_t)$, respectively. The MDP tuples are described as follows:
\subsubsection{State} In each time step, $s_t \in \mathcal{S}$ indicates the current status of O-RAN. This status includes information such as the QoS that each slice can achieve, $Q_l$, the density of UEs in each slice, $N_l(t)$, and the previous action taken for resource allocation, $a_{t-1}$. Thus, at time $t$, the intelligent agent's observation can be expressed as $s_t = \{Q_l,N_l,a_{t-1}\mid \forall l\in \mathcal{L}\}$.
\subsubsection{Action} The vector $a_t \in \mathcal{A}$ represents the number of resources required for the O-RAN slices and UEs. Therefore, in each time step $t$, the agent uses its policy to determine the appropriate action to take as $a_t = \{\boldsymbol{e},\boldsymbol{b}\}$.

\subsubsection{Reward} The value of $r_t$ is determined by summation of the probability of SLA violation as described in equation \eqref{opt1}. This probability is influenced by the incoming traffic for each slice and the radio conditions of the UEs that are connected to the network. The desired reward value is defined as: 
   \begin{equation}
        r_t = \sum_{l=1}^{\abs{\mathcal{L}}}\Big( \mathbbm{P}(\abs{Q_l(\boldsymbol{e},\boldsymbol{b})-\lambda_l} \leq  \epsilon_l) \Big).
    \end{equation}
Therefore, the procedure aims to optimize the utilization of the available bandwidth to satisfy the QoS requirements of all the network slices. After this, the MDP model that has been defined can be explored through the use of a DRL approach. In a DRL framework, the key objective of the agent is to determine the best policy $\pi^*(a_t|s_t;\theta_p)$, which is a mapping from the state space to the action space. This optimal policy aims to maximize the expected average discounted reward $\mathbb{E}{\pi}[R(t)]$, where $R(t) = \sum_{i=0}^{\infty}\gamma^i r_{i,t}$ is the sum of the discounted rewards $r_{i,t}$, and $\gamma$ is the discount factor used for future rewards. Given a policy $\pi$, the state-value and action-value functions are defined as $V(s_{t}) = \mathbb{E}_{\pi}[R(t)\mid s_{t}]$ and $Q(s_{t},a_{t}) = \mathbb{E}_{\pi}[R(t)\mid s_{t},a_{t}]$, respectively. 

A single DRL agent in the near-RT RIC monitors and controls an O-RAN network with disaggregated modules. However, training a single DRL agent while considering the heterogeneity of UEs' services, as discussed in~\cite{polese2022colo,hammami2022policy}, can be time-consuming and may result in a suboptimal solution due to a potential lack of exploration scenarios. To address this, leveraging the opportunities provided by O-RAN, employing multiple DRL agents located in O-DU locations and distributed across the network can expedite and stabilize the training process by utilizing the collective experience and different exploration policies of all the disaggregated modules. This approach takes advantage of the rich source of knowledge and diversity available across the network components to stabilize training. For instance, since various O-DUs in an O-RAN system are deployed in different locations across the network, they experience diverse network conditions, including network traffic and QoS requirements. Sharing experiences among the distributed DRL agents enables the exploration of a wider range of scenarios, enhancing the generality of resource allocation. 
Despite potential communication costs, the benefits of utilizing distributed DRL agents in training outweigh these costs. Moreover, an actor-critic approach offers advantages in efficient exploration-exploitation, sample efficiency, flexibility, adaptability, and online learning, making it a popular choice for solving complex RL problems. Hence, utilizing multiple actors across the O-RAN network with a single global critic provides advantages such as optimality of solutions, access to diverse experiences of DUs, generalizability, and scalability.\vspace{-0.cm}

To learn from experiences generated by different exploration policies, off-policy methods are required. Since the states in this paper are continuous, a Soft Actor-critic (SAC) algorithm has been used. SAC is known to be more effective in high-dimensional or continuous environments due to its entropy regularization term that prevents the policy from converging too quickly to a suboptimal solution~\cite{haarnoja2018soft}. Compared to other off-policy methods, SAC strikes a balance between exploration and exploitation and achieves better sample efficiency, requiring fewer environmental interactions. This off-policy nature of SAC also makes it sample-efficient while maintaining a stable learning process. 
Here, distributed intelligent actors are located in the O-DU modules across the network and utilize the actor-critic technique to train an ideal policy that assigns resources in a way that maximizes the long-term reward. To do that, a global critic is considered in the O-CU module location, in the RIC module according to Fig. \ref{sys_graph}. The policy network parameterized by $\boldsymbol{\theta}_p$ is updated using the gradient defined as follows with $\kappa$ random samples transitions:
\begin{align}\label{pupdate} 
    \nabla_{\theta_p}J(\pi_\theta) =  \mathbb{E}_{\kappa,\pi}\big[ \nabla_{\theta_p} \log(\pi_{\theta_p}(a|s)) &\big(-\beta \log(\pi_{\theta_p}(a|s)) \nonumber\\
    &+ Q(s,a;\theta_v)\big)\big],
\end{align}
where $\beta$ is a temperature parameter that determines the balance between maximizing entropy and reward. 
Also, the value network parameterized by $\boldsymbol{\theta}_v$ will be updated by minimizing the following loss as:\vspace{-0.2cm}
\begin{align}\label{vupdate} 
    \min_{\theta_v} \mathbb{E}_{\kappa,\pi} \bigg(y_i-Q_{\pi_i}(s_{i},a_{i};\theta_v)\bigg)^2, \vspace{-0.1cm}
\end{align}
where $y_i = r_i + \gamma Q_{\pi_{i+1}}(s_{i+1},a_{i+1};\theta_v)-\beta \log(\pi_{\theta_p}(a|s)$. 
While leveraging collective experience information from different locations can be beneficial for training, a challenge of such a model is that not all information may be useful. Actually, treating all agents uniformly at all times can overwhelm the critic with unnecessary updates from less informative scenarios, potentially decreasing the convergence speed of training. For instance, certain locations may experience simple scenarios most of the time, providing limited useful information for the critic. Therefore, equal treatment of all experience information may result in the critic spending more time on trivial information, and less time on useful information, impacting training efficiency. Hence, inspired by \cite{foerster2018counterfactual,iqbal2019actor}, we employ an attention-based strategy for cooperation between distributed agents to solve the problem \eqref{opt1}. This attention-based strategy assumes the role of an intelligent leader, guiding the critic network to acquire valuable insights from challenging locations.  \vspace{-0.3cm}
\begin{figure}[t!]
  \centering
    \includegraphics[width=0.74\columnwidth]
    {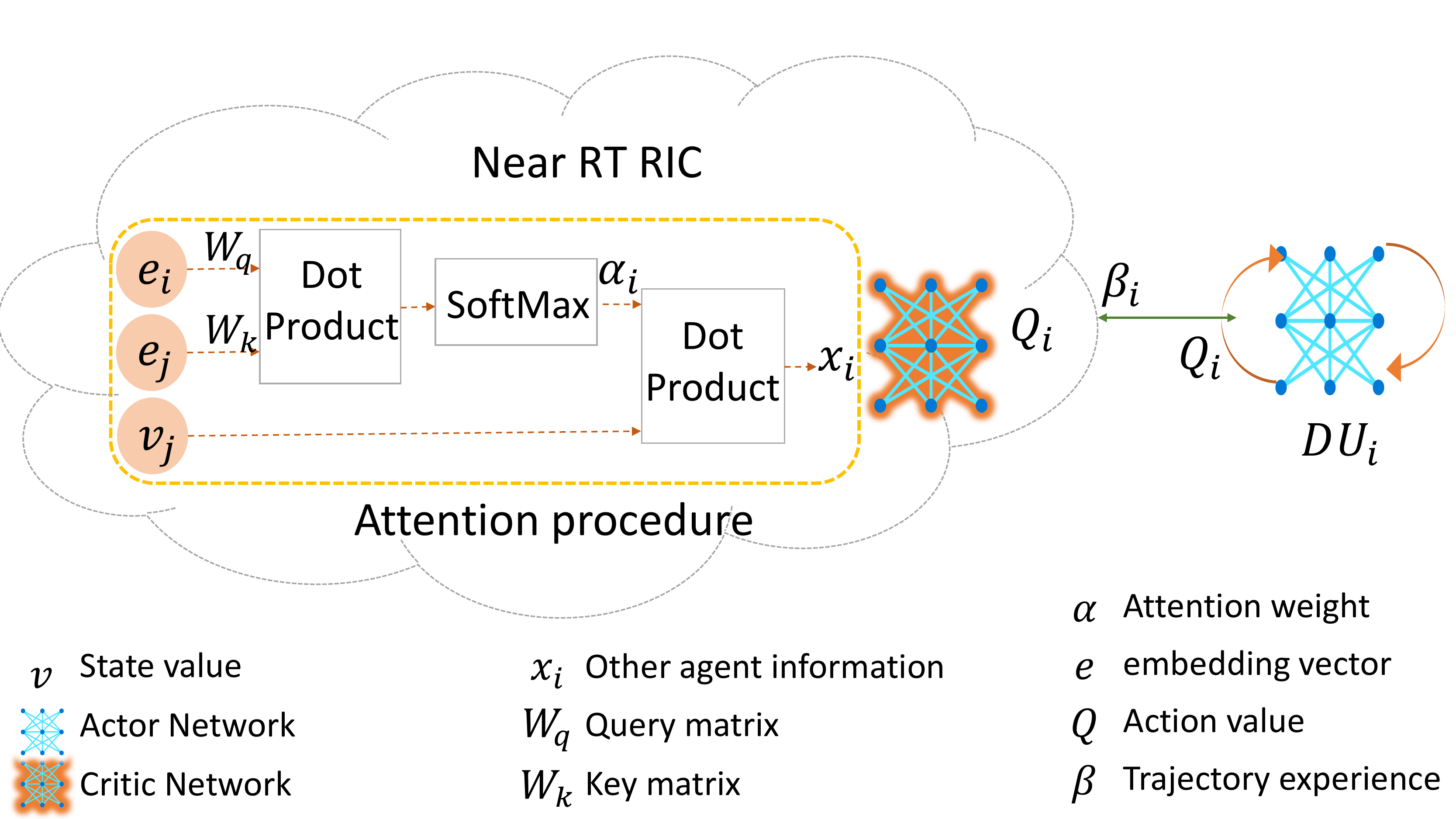}\vspace{-0.1cm}
    \caption{\small The proposed attention-based DRL in Open RAN network. 
    }\vspace{-0.1cm}
    \label{sys_proposed}
\end{figure}
\section{Attention-based distributed DRL strategy}\label{ADRL}

Treating all experience information equally may potentially decrease the performance of the training process, despite the fact that not all information may be useful. To address this challenge, we propose an attention-based distributed DRL (ADRL) strategy for cooperation between distributed agents for O-RAN slicing issues. This approach allows the critic network to selectively focus on the most relevant and valuable information from challenging network situations during training, leading to improved learning and performance. By leveraging the attention mechanism, our proposed solution aims to enhance the training process and address the issue of potentially sub-optimal solutions due to irrelevant or limited information. Our proposed multi-agent learning approach aims to stabilize the training of individual agents by selectively attending to information from other agents. This methodology closely follows the established paradigm of centrally training agents in order to address challenges arising from dynamic environments, while simultaneously implementing distributed policy execution.\vspace{-0.2cm} 

\subsection{The proposed ADRL approach}
The proposed approach employs a value-based attention mechanism to enable the critic to assign varying levels of importance to different observation data according to Fig. \ref{sys_proposed}. Specifically, each agent requests information from other agents regarding their observations and actions, which is subsequently incorporated into the value function estimate. Through the use of an attention network inspired by \cite{iqbal2019actor}, the $Q$ value of each agent leverages both its own information as well as relevant information from other agents, as formulated by the following equation. \vspace{-0.2cm}
\begin{align}\label{q_calc}
    Q_i(s,a) = C(e_i,x_i),
\end{align}
where $C$ is a neural network that calculates the $Q_i$ value and will be trained through the system training process. Also, $e_i$ represents an embedding value of agent $i$ that is calculated with an encoder neural network $g_i$ as an embedding function over $s_i$ and $a_i$ as $e_i = g_i(s_i,a_i)$. The $x_i$ represents the information of other agents which is a weighted sum of other agents' value according to the following equation. 
\begin{align}\label{x_calc}
    x_i = \sum_{j\neq i}\alpha_j \Tilde{v}_j,
\end{align}

$\Tilde{v}_j$ shows a variant of $v_j$, where leaky ReLU is utilized as an element-wise nonlinearity. In addition, $\alpha_j$ shows an attention weight that compares embedding values $e_i$ with $e_j$ using a bilinear mapping as the query-key system. Then, the attention weight $\alpha_i$ will be measured by passing the similarity value into a softmax as follows:\vspace{-0.1cm}
\begin{align}\label{attention_weight}
    \alpha_j = \text{softmax} \big(e_j^T W_k^T W_q e_i\big),
\end{align}
where $W_k$ transforms $e_j$ into a "key" value, and $W_q$ transforms $e_i$ into a "query" value.  
\begin{algorithm}[t!]
\SetAlgoLined
\textbf{Input}: $N_t$,\,\,$N_m$,\,\,$N_e$\,\,$\theta_{p,i},\forall i \in [0,N_m]$,\,\,$\theta_{A}$,\,\,$\theta_v$.    \\
\For{iteration $t=1:N_t$}{
\For{actor $i=1:N_m$}{
\For{evaluation $e = 1 : N_e$}{
$r_i = \text{evaluate}(\pi_{p,i})$.\\
$\mathcal{B}\gets \langle s_t,a_t,s_{t+1},r_t \rangle $.
}
}
Update the $\theta_{A}$ of the attention network by calculating $Q_i$ using \eqref{q_calc}, \eqref{x_calc}, \eqref{attention_weight} and batch of $\mathcal{B}$. \\
Update $\theta_v$ using \eqref{vupdate} and all the $Q_i$ values.\\
Update $\theta_p$ using \eqref{pupdate}.\\
 \If{$\theta_p$ by \eqref{pupdate} is converged}{
 Break.
 }
}
\textbf{Output}: $\theta_{p,i},\forall i \in [0,N_m]$,\,\,$\theta_v$. \\
\caption{The ADRL algorithm}\vspace{-0.1cm}
\label{alg1}
\end{algorithm}\vspace{-0cm}

Algorithm \ref{alg1} summarizes the ADRL strategy to solve the optimization problem in \eqref{opt1}-\eqref{opt1_e}. The input variables of the algorithm are the number of iterations $N_t$, the number of distributed actors $N_m$, the number of evaluations $N_e$, and the distributed actors' network weights as $\theta_{p,i}$ which are initialized with random weights. Also, the critic network and attention network weights as $\theta_v$ and $\theta_A$ are initialized randomly. 
The algorithm proceeds to output the trained policy of each distributed DRL agent. In each iteration loop $t$, the distributed actors have $N_e$ number of evaluations to have enough interactions with their environment. Then, each distributed actor's experience is stored in the replay buffer $\mathcal{B}$. Next, by considering part of the replay buffer data, in the attention network, the attention weights and other agents' information are calculated using \eqref{attention_weight} and \eqref{x_calc}, respectively. Then, the $Q$ value can be measured using \eqref{q_calc} to update the critic network $\theta_v$ using \eqref{vupdate}. In this step, the global critic network updates all the distributed actor networks by using \eqref{pupdate}. The algorithm terminates once the distributed DRL actors' policy network is converged or after $N_t$ maximum iteration time. \vspace{-0.2cm}

\begin{table}[t!] 
	\footnotesize
	\centering
	\caption{\vspace*{-0cm} Simulation parameters} \vspace{-0.2cm}
	\begin{tabular}{|>{\centering\arraybackslash}m{2.4cm}|>{\centering\arraybackslash}m{1.7cm}|>{\centering\arraybackslash}m{1.2cm}|>{\centering\arraybackslash}m{1.7cm}|}
		\hline
		\bf{Parameter} &\bf{Value } & \bf{Parameter} &\bf{Value }\\
		\hline
		Subcarrier spacing & $15$ kHz & $h$ & Rayleigh fading channel \\
		\hline
		Total bandwidth/DU & \{$10$, $20$\}MHz  & $\sigma^2$ & $-173$ dBm \\
		\hline
		RB bandwidth/DU  & $200$ kHz & $N_t$ & $100$ \\
		\hline
		$p_u$ & $56$ dBm & $N_m$ & $6$ \\
		\hline
		$K$ per DU & \{$50$, $200$\} & $N_e$ & $10$ \\
		\hline
		$N$ per DU & $50$ & batch size & $128$ \\
		\hline
	\end{tabular}\label{param} \vspace{-0.2cm}
\end{table}

\section{simulation results}\label{sim}

The simulation involves an O-RAN design with three slices (eMBB, MTC, URLLC) serving 50 users per DUs, which are uniformly distributed and randomly located across the network. 
We consider the fact that the distribution of demand services across the network is not uniform, 
resulting in each DU being exposed to varying levels of UE traffic and prioritizing different services for each DU. Here, similar to \cite{cheng2022reinforcement} we model the traffic based on the UEs density and UEs have mobility into the cell range. 
To implement the DRL approach, we consider an actor-critic approach using the Pytorch library with three fully-connected layers with $128$, $256$, and $256$ neurons for actor and critic networks and a \textit{tanh} activation function. For all the models, the learning rate $10^{-4}$ and \textit{Adam} optimizer are considered. 
For the attention network, we implement one fully-connected layer with $128$ neurons for the encoder network, and two fully-connected layers with $256,256$ neurons to measure the $Q$ value. 
Table \ref{param} summarizes other simulation parameters. In this simulation, we consider a distributed-DRL approach in which the actors are distributed across the network, and a global critic is located in the RIC module as a baseline to compare with our proposed ADRL method. 

\begin{figure}[t!]
  \centering
    \includegraphics[width=7.6cm]{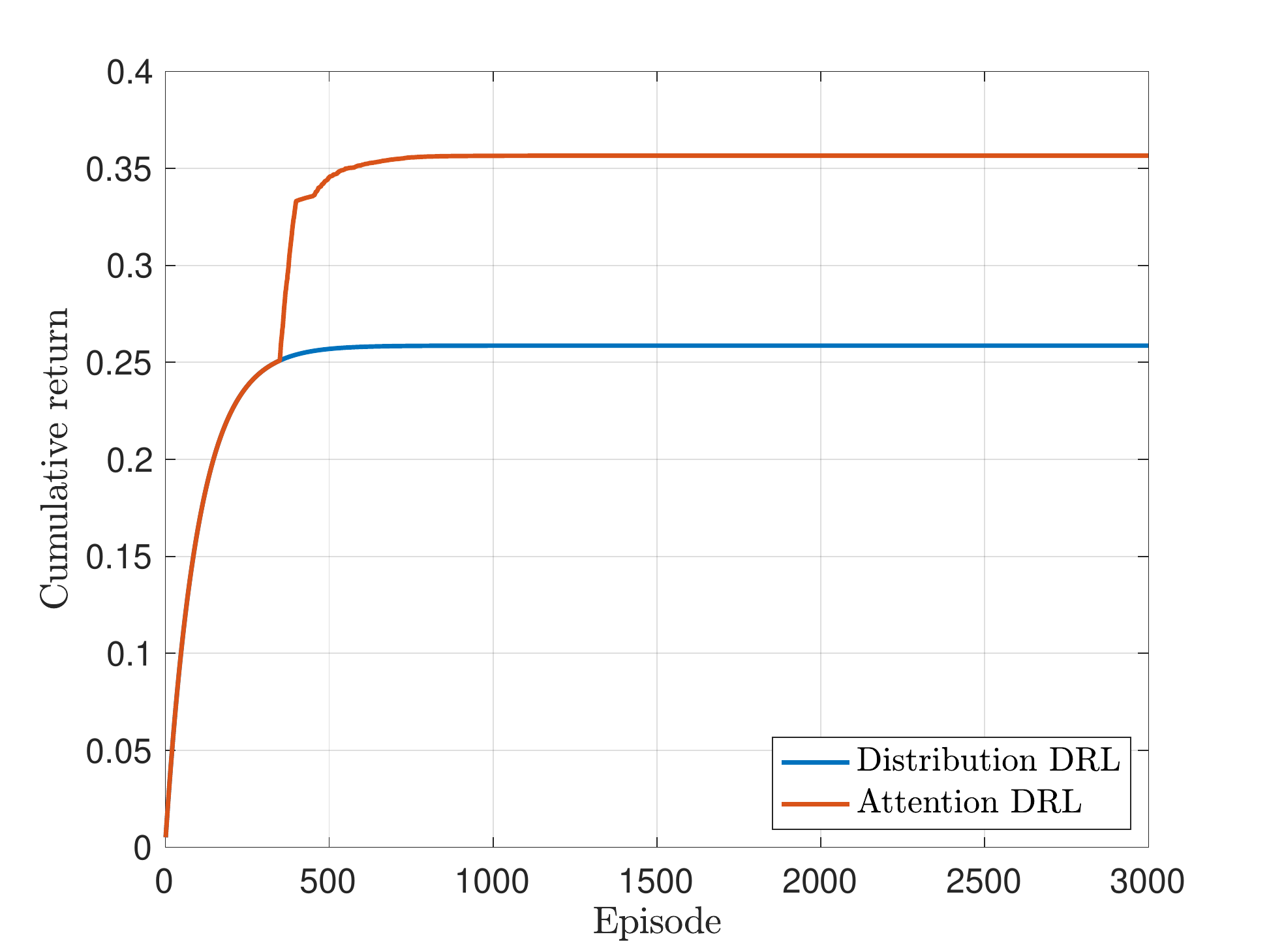}\vspace{-0.2cm}
    \caption{\small Performance comparison of attention-DRL and distributed-DRL algorithm.}\vspace{-0.2cm}
    \label{cum_return}
\end{figure}
Figure \ref{cum_return} compares the performance of the proposed attention-DRL algorithm with the distributed-DRL algorithm as a baseline method in a wireless environment. The cumulative rewards were measured with $\gamma=0.99$ and the results were averaged over a sufficient number of runs. Figure \ref{cum_return} displays the results in each episode for comparison. The results reveal that the attention-DRL approach can provide up to a $32.8\%$ greater final return value than the other DRL method, demonstrating the efficacy of the proposed ADRL algorithm in the wireless environment. At the early episodes, the proposed method has almost equal return value with the distributed-DRL method as the attention network requires a couple of episodes to be trained. Then, the attention network shows its effect on the final return value of the attention-DRL method, 
and the proposed method, which utilizes useful information, outperforms the distributed-DRL approach. 

\begin{figure}[t!]
  \centering
    \includegraphics[width=7.6cm]{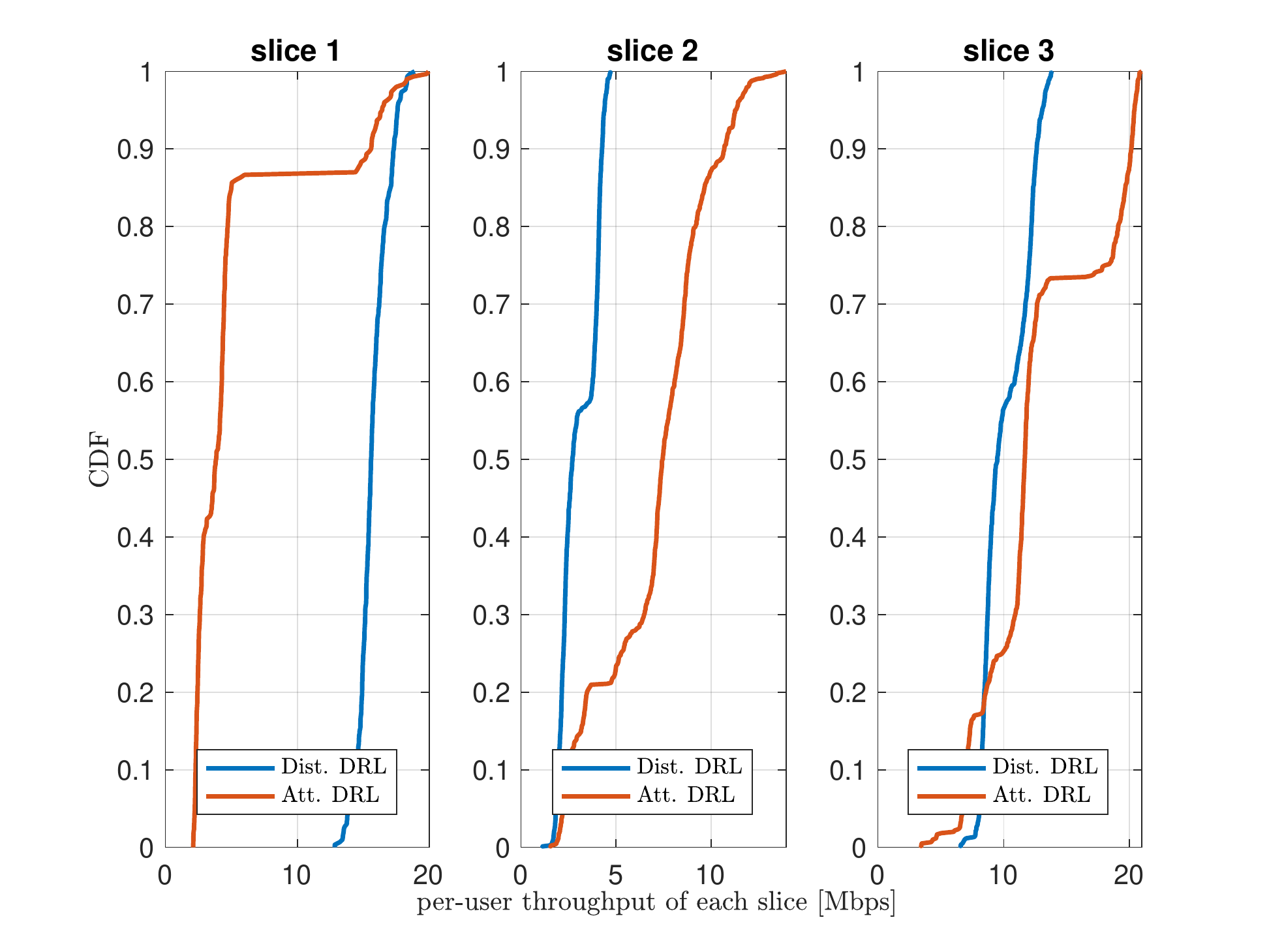}\vspace{-0.2cm}
    \caption{\small CDF of achieved throughput per users through attention-DRL and distributed-DRL training process in a low BW scenario.}\vspace{-0.2cm}
    \label{cdf_UE}
\end{figure}
\begin{figure}[t!]
  \centering
    \includegraphics[width=7.6cm]{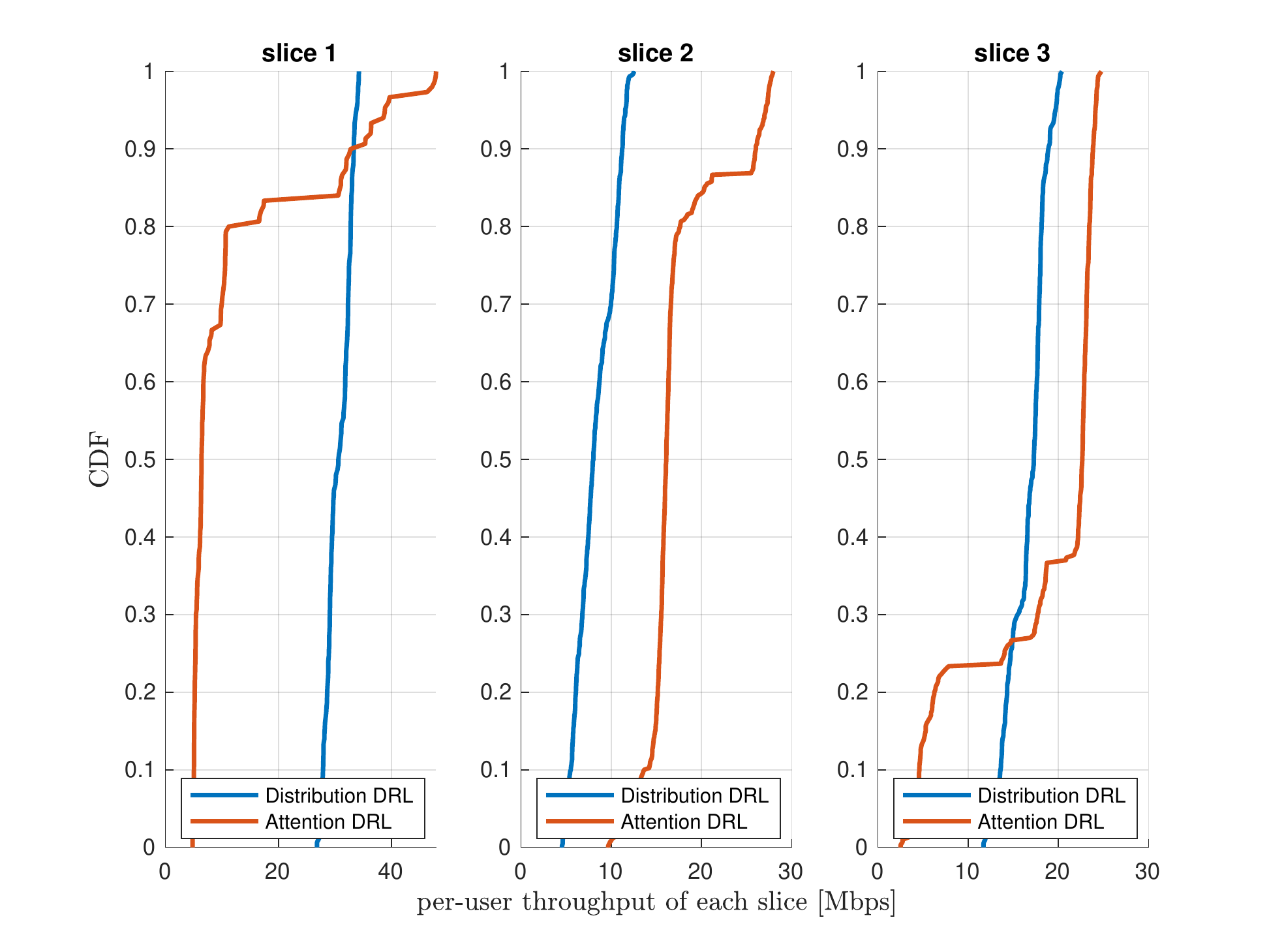}\vspace{-0.2cm}
    \caption{\small CDF of achieved throughput per users through attention-DRL and distributed-DRL training process in a high BW scenario.}\vspace{-0.2cm}
    \label{cdf_UE_h}
\end{figure}

Figures \ref{cdf_UE} and \ref{cdf_UE_h} display the Cumulative Distribution Functions (CDFs) of the per-user throughput in each slice of the simulation O-RAN environment during the attention-DRL and distributed-DRL training processes in a low BW ($K = 50$ RBs) and high BW ($K = 200$ RBs) regime scenarios, respectively. 
The three slices considered are eMBB, MTC, and URLLC, in that order. Each slice meets the service demands by considering a distinct QoS for specific services. For example, the eMBB slice uses the average data rate as a QoS metric to guarantee a stable service for connected UEs. Similarly, the MTC slice uses capacity as a QoS parameter to ensure high connection density. In addition, the URLLC slice considers an exponential random variable with an average size of $10$ Kb as packet length and selects maximum delay as a QoS criterion to ensure the lowest possible value for maximum latency in these delay-sensitive services. 
The results displayed were obtained during training iterations by distributing users to the slices in each DU non-uniformly. As observed in Figures \ref{cdf_UE} and \ref{cdf_UE_h}, the network users adhere to the QoS demand of each slice. Additionally, the results demonstrate that the proposed Algorithm \ref{alg1} was effective in network users' throughput compared to the baseline in both low BW and high BW scenarios.

\begin{figure}[t!]
  \centering
    \includegraphics[width=7.6cm]{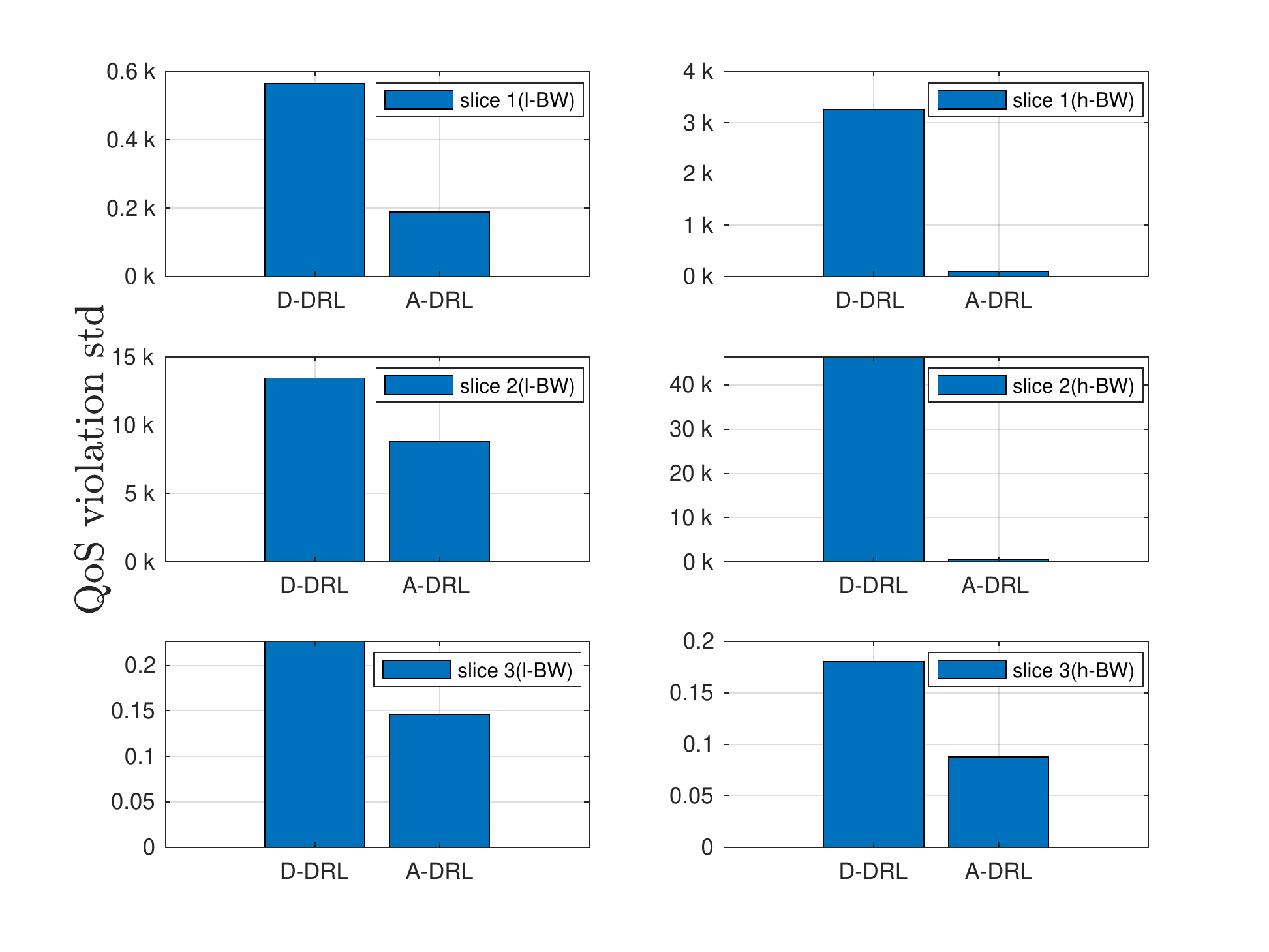}\vspace{-0.2cm}
    \caption{\small QoS violation STD of each slice for attention-DRL and distributed-DRL methods in a low and high BW scenarios}\vspace{-0.2cm}
    \label{std_q}
\end{figure}
Figure \ref{std_q} illustrates a comparison of the standard deviation (std) values for each slice's QoS violation in attention-DRL and distributed-DRL methods for low and high BW scenarios. As mentioned previously, the objective of this study was to manage and prevent QoS violations of various services in the network. This figure demonstrates that the proposed attention-DRL method can effectively reduce SLA violations in the system and outperforms the baseline in both low and high BW situations. The results presented in Figure \ref{std_q} indicate that the incorporation of attention networks and the utilization of distributed actors' valuable information are beneficial for network management.  \vspace{-0.2cm}

\section{Conclusion}\label{conclusion}

We propose a novel attention-DRL framework for the efficient allocation of shared radio resources among O-RAN slices. Our approach utilizes distributed experiences to develop a robust and general policy that performs well in various traffic situations. The attention-based strategy differentiates crucial information, leading to better results and preventing misleading data. Our simulation results demonstrate up to a $32.8\%$ improvement in maximum rewards compared to the baseline distributed-DRL method, highlighting the importance of relevant experiences and generalization in policy training in dynamic wireless networks. The proposed algorithm is also efficient in the presence of wireless bandwidth constraints.\vspace{-0.1cm}

\def\baselinestretch{0.9}
\bibliographystyle{IEEEbib}
\bibliography{Main}

\begin{thebibliography}{10}

\bibitem{li2020ran}
C.~{Li} and A.~{Akman},
\newblock ``{O-RAN} use cases and deployment scenarios. towards open and smart
  ran,''
\newblock {\em White paper}, 2020.

\bibitem{3gpprelease16}
3GPP TS~23.501 version 16.6.0 Release~16,
\newblock ``{5G}; system architecture for the {5G} system {(5GS)},''
\newblock {\em Tech. Spec.}, , no. 3, 2020.

\bibitem{popovski20185g}
P.~{Popovski}, K.~{Trillingsgaard}, O.~{Simeone}, and G.~{Durisi},
\newblock ``{5G} wireless network slicing for {eMBB}, {URLLC}, and {mMTC}: A
  communication-theoretic view,''
\newblock {\em IEEE Access}, vol. 6, pp. 55765--55779, 2018.

\bibitem{yang2020data}
H.~{Yang}, A.~{Yu}, J.~{Zhang}, J.~{Nan}, B.~{Bao}, Q.~{Yao}, and M.~{Cheriet},
\newblock ``Data-driven network slicing from core to {RAN} for {5G}
  broadcasting services,''
\newblock {\em IEEE Transactions on Broadcasting}, vol. 67, no. 1, pp. 23--32,
  2020.

\bibitem{3gpp2017study}
3GPP TR~38.912 version 14.1.0 Release~14,
\newblock ``Study on new radio access technology: Radio access architecture and
  interfaces,''
\newblock {\em Tech. Rep}, , no. 3, 2017.

\bibitem{oranslice2020}
O-RAN Working~Group 1,
\newblock ``Study on o-ran slicing-v2.00,''
\newblock {\em O-RAN.WG1.Study-on-O-{RAN}-Slicing-v02.00 Technical
  Specification}, April 2020.

\bibitem{polese2022understanding}
M.~{Polese}, L.~{Bonati}, S.~{D'Oro}, S.~{Basagni}, and T.~{Melodia},
\newblock ``Understanding {O-RAN}: Architecture, interfaces, algorithms,
  security, and research challenges,''
\newblock {\em arXiv preprint arXiv:2202.01032}, 2022.

\bibitem{thaliath2022predictive}
J.~{Thaliath}, S.~{Niknam}, S.~{Singh}, R.~{Banerji}, N.~{Saxena}, HS.
  {Dhillon}, JH. {Reed}, AK. {Bashir}, A.~{Bhat}, and A.~{Roy},
\newblock ``Predictive closed-loop service automation in o-ran based network
  slicing,''
\newblock {\em IEEE Communications Standards Magazine}, vol. 6, no. 3, pp.
  8--14, 2022.

\bibitem{cheng2022reinforcement}
NF. {Cheng}, T.~{Pamuklu}, and M.~{Erol-Kantarci},
\newblock ``Reinforcement learning based resource allocation for network slices
  in {O-RAN} midhaul,''
\newblock pp. 140--145, 2023.

\bibitem{zhang2022team}
H.~{Zhang}, H.~{Zhou}, and M.~{Erol-Kantarci},
\newblock ``Team learning-based resource allocation for open radio access
  network {(O-RAN)},''
\newblock in {\em ICC 2022-IEEE International Conference on Communications}.
  IEEE, 2022, pp. 4938--4943.

\bibitem{zhang2022federated}
H.~{Zhang}, H.~{Zhou}, and M.~{Erol-Kantarci},
\newblock ``Federated deep reinforcement learning for resource allocation in
  {O-RAN} slicing,''
\newblock in {\em GLOBECOM 2022-2022 IEEE Global Communications Conference}.
  IEEE, 2022, pp. 958--963.

\bibitem{lotfi2022evolutionary}
F.~{Lotfi}, O.~{Semiari}, and F.~{Afghah},
\newblock ``Evolutionary deep reinforcement learning for dynamic slice
  management in {O-RAN},''
\newblock in {\em 2022 IEEE Globecom Workshops (GC Wkshps)}. IEEE, 2022, pp.
  227--232.

\bibitem{polese2022colo}
M.~{Polese}, L.~{Bonati}, S.~{D’Oro}, S.~{Basagni}, and T.~{Melodia},
\newblock ``{ColO-RAN}: Developing machine learning-based xapps for open ran
  closed-loop control on programmable experimental platforms,''
\newblock {\em IEEE Transactions on Mobile Computing}, 2022.

\bibitem{hammami2022policy}
N.~{Hammami} and K.~{Nguyen},
\newblock ``On-policy vs. off-policy deep reinforcement learning for resource
  allocation in open radio access network,''
\newblock in {\em 2022 IEEE Wireless Communications and Networking Conference
  (WCNC)}. IEEE, 2022, pp. 1461--1466.

\bibitem{haarnoja2018soft}
T.~{Haarnoja}, A.~{Zhou}, P.~{Abbeel}, and S.~{Levine},
\newblock ``Soft actor-critic: Off-policy maximum entropy deep reinforcement
  learning with a stochastic actor,''
\newblock in {\em International conference on machine learning}. PMLR, 2018,
  pp. 1861--1870.

\bibitem{foerster2018counterfactual}
J.~{Foerster}, G.~{Farquhar}, T.~{Afouras}, N.~{Nardelli}, and S.~{Whiteson},
\newblock ``Counterfactual multi-agent policy gradients,''
\newblock in {\em Proceedings of the AAAI conference on artificial
  intelligence}, 2018, vol.~32.

\bibitem{iqbal2019actor}
S.~{Iqbal} and F.~{Sha},
\newblock ``Actor-attention-critic for multi-agent reinforcement learning,''
\newblock in {\em International conference on machine learning}. PMLR, 2019,
  pp. 2961--2970.

\end{thebibliography}
\end{document}